\documentclass[12pt]{article}
\usepackage{graphicx}
\usepackage{amssymb} 
\usepackage{amsmath}
\usepackage{hyperref}

\newcommand\pubnumber{DPF2013-139}
\newcommand\pubdate{\today}

\def\rpa{$R$-parity}
\def\uone{U(1)}
\def\HH{\mathcal{H}}
\def\OO{\mathcal{O}}
\def\nn{\nonumber}
\def\nmu{n_\mu}

\def\ucsc{Santa Cruz Institute for Particle Physics and\\
 Department of Physics, 
  University of California, Santa Cruz CA 95064}

\def\Title#1{\begin{center} {\Large #1 } \end{center}}
\def\Author#1{\begin{center}{ \sc #1} \end{center}}
\def\Address#1{\begin{center}{ \it #1} \end{center}}

\newcommand\pubblock{\rightline{\begin{tabular}{l} \pubnumber\\
         \pubdate  \end{tabular}}}
\newenvironment{Abstract}{\begin{quotation}  }{\end{quotation}}
\newenvironment{Presented}{\begin{quotation} \begin{center} 
             PRESENTED AT\end{center}\bigskip 
      \begin{center}\begin{large}}{\end{large}\end{center} \end{quotation}}
\def\Acknowledgments{\bigskip  \bigskip \begin{center} \begin{large}
             \bf ACKNOWLEDGMENTS \end{large}\end{center}}

\def\beq{\begin{equation}}
\def\eeq#1{\label{#1}\end{equation}}
\def\eeqn{\end{equation}}

\def\beqa{\begin{eqnarray}}
\def\eeqa#1{\label{#1}\end{eqnarray}}
\def\eeqan{\end{eqnarray}}

\let\bar=\overbar

\def\vev#1{\langle #1 \rangle}

\def\Dslash{\not{\hbox{\kern-4pt $D$}}}
\def\dslash{\not{\hbox{\kern-2pt $\del$}}}

\def\msb{{\bar{\ssstyle M \kern -1pt S}}}

\def\eps{\epsilon}

\begin{document}
\begin{titlepage}
\pubblock

\vfill
\Title{Low-energy R-parity violating SUSY  with horizontal\\ flavor symmetries}
\vfill
\Author{ Angelo Monteux}
\Address{\ucsc}
\vfill
\begin{Abstract}
In this talk, I will present the general structure of RPV couplings when a Froggatt-Nielsen horizontal symmetry is responsible for the flavor structure of both the SM and the MSSM. 
For sub-TeV ({\it natural}) SUSY, lepton number must be an accidental symmetry, while low-energy SUSY is still allowed by baryonic RPV, which lowers the MET signature of superparticles decays. The largest RPV coupling involves the stop, and it is constrained between $10^{-3}$ (from FCNCs) and $10^{-9}$ (from LHC searches).

\end{Abstract}
\vfill
\begin{Presented}
DPF 2013\\
The Meeting of the American Physical Society\\
Division of Particles and Fields\\
Santa Cruz, California, August 13--17, 2013\\
\end{Presented}
\vfill
\end{titlepage}
\def\thefootnote{\fnsymbol{footnote}}
\setcounter{footnote}{0}

\section{Introduction}
While Supersymmetry (SUSY) is a well-motivated extension of the Standard Model (SM), current LHC searches put stringent bounds on the masses of the superparticles; because they are strongly interacting, gluinos should be easily produced at the LHC, and current bounds are at 1.4 TeV when assuming simplified decay chains to squarks and neutralinos.\footnote{Limits on stops, more relevant for the naturalness of the weak scale, are  at about 600 GeV, with a strong dependence on the stop branching ratios; regardless, the gluino enters the stop RGE equation, so that a light stop and a heavy gluino would result in a fine-tuning of the stop mass.} In most searches, $R$-parity, a symmetry under which superparticles have opposite charges with respect to SM fields, is assumed.\footnote{$R$-parity can be expressed as $R_p=(-1)^{2S+3B+L}$.} Hence, the lightest supersymmetric partner (LSP) is stable and exits the detector, showing up as missing energy (and providing a dark matter candidate). Without \rpa, the LSP can decay to SM particles, possibly charged and/or colored, and the missing energy signature of a supersymmetric event is greatly reduced.

The strict bounds on \rpa\, conserving SUSY might be hinting to the possibility of \rpa-violating (RPV) SUSY; in particular, one should think of why \rpa\, was introduced and if it is superfluous. 

\rpa\, is introduced to forbid renormalizable operators which violate both lepton number ($L$) and baryon number ($B$):
\beq 
W_{RPV}=\bar \mu_i L_i\phi_u+\lambda_{ijk} L_iL_j\bar\ell_k +\lambda'_{ijk}L_iQ_j\bar d_k+\lambda_{ijk}''\bar u_i\bar d_j\bar d_k\,.
\eeqn
The first three operators break $L$, while the last breaks $B$. A combination of both types is needed for the proton to decay to lighter particles.

Still, there are dimension-5 operators allowed by \rpa, as $\frac{1}{M_P}{Q}_i{Q}_j{Q}_k{L}_l$, that would produce proton decay at dangerous rates if a particular flavor structure is not assumed. It is then reasonable to try to understand the flavor structure of the (MS)SM and see if imposing \rpa\ was after all not necessary (a dark matter candidate can be given by the gravitino, if sufficiently long-lived, or an axion/axino mixture).

In ref. \cite{Monteux:2013mna}, we used a  Froggatt-Nielsen mechanism of horizontal symmetries \cite{Froggatt:1978nt,Leurer:1992wg} to show that the RPV couplings hierarchies are fixed by the observed SM mass hierarchies and mixings (for similar analysis, see \cite{Florez:2013mxa,Joshipura:2000sn}). Furthermore, both low-energy and collider phenomenology  were studied for this model, leading to the prediction of an accidental lepton number conservation (at the renormalizable level) and a specific range allowed for the largest baryonic RPV coupling, $\lambda''_{323}$.

In section \ref{horiz} we review the framework of horizontal symmetries and how it can explain the flavor structure of the MSSM, including the RPV coefficients. Textures for these couplings will be uniquely determined by the measured SM flavor structure. In section \ref{pheno}, we present the phenomenological limits for the coefficients, and the applicable bounds coming from collider searches. We summarize in section \ref{conclusions} and give an outlook for the next LHC searches.

\section{Textures from horizontal symmetries}\label{horiz}
The Froggatt-Nielsen framework \cite{Froggatt:1978nt} assumes a new horizontal symmetry, $\uone_\HH$, under which the SM fields have generation-dependent charges. The high-energy theory is invariant under $\uone_\HH$, which is broken when  a flavon field $S$, with charge $-1$, aquires a vev $\vev{S}$. In the low-energy theory, heavy fields that have been integrated out generate effective operators proportional to a spurion $\eps=\frac{\vev S}M$, where $M$ is the symmetry-breaking scale. Only terms that are invariant under the symmetry are allowed in the superpotential (including different powers of the spurion). A Yukawa coupling $Y^d_{ij}\phi_dQ_i\bar d_j$ is rewritten as
\beq
\eps^{m_{ij}}\phi_dQ_i\bar d_j,\qquad m_{ij}=\HH[\phi_d]+\HH[Q_i]+\HH[\bar d_j]-r\,,
\eeqn
where $r=0$ for a non $R$-symmetry and $r=2q_\theta$ if $\uone_\HH$ is an $R$-symmetry. Unknown $\OO(1)$ factors have been neglected in front of the operator; one expects that the hierarchies are generated by the different charges and not by order one parameters, and whenever this is not possible a tuning is present. 
Operators with negative or fractional powers of $\eps$ are not allowed in the superpotential.
Using the notation $\HH[\Phi_i]=\Phi_i$, the Yukawas are
\beq
Y_{ij}^a=\eps^{\phi_a+Q_i+a_j-r}, \qquad a=u,d;\ i,j=1,2,3.
\eeqn
Therefore, the masses and mixings are also written in terms of the horizontal charges:
\beq
 \frac{m^a_i}{m^a_j}\sim\eps^{Q_i+a_i-Q_j-a_j},  \qquad  |V_{ij}|\sim\eps^{|Q_i-Q_j|}.
\eeqn
Similar expression holds for the leptons. Given the experimental values for masses and mixings, and taking the magnitude of the spurion equal to the largest of the ``small'' SM parameters, $\eps=\sin\theta_C$, the charge differences $\Phi_{ij}\equiv \Phi_i-\Phi_j$ are uniquely set as follows:
\beq\nn
\begin{array}{|ccc|ccc|ccc|ccc|ccc|}\hline
Q_{12}&Q_{13}&Q_{23}&d_{12}&d_{13}&d_{23}&u_{12}&u_{13}&u_{23}&L_{12}&L_{13}&L_{23}&\ell_{12}&\ell_{13}&\ell_{23}\\\hline
1&3&2&  1&2&1&  3&5&2&0&0&0&4&6&2\\\hline
\end{array}
\eeqn
With  $17$ charges and only $13$ independent equations, the solutions depend on the choices of 4 independent variables, which can be taken as $\{Q_3, u_3, d_3, L_3\}$.\footnote{
If $\nmu=\phi_u+\phi_d<0$, the  $\mu$ term $\mu\phi_u\phi_d$  is generated by a Giudice-Masiero-like term Kh\"aler correction $\delta K=X\phi_u\phi_d\left(\frac{S^*}M\right)^{-n_\mu}$, resulting in $\mu=m_{3/2}\eps^{|\nmu|}$.
The $\mu$ term is automatically suppressed with respect to the SUSY breaking scale. If the mixed $\uone_\HH$-gauge anomalies are cancelled by a universal axion, we have $\nmu\sim-1$. As pointed out in \cite{Dine:2012mf},  non-universal anomalies are more common in string-theory motivated models, so the requirement of anomaly cancellation is not considered.
}

\subsection{\rpa\, violation}
In the presence of a horizontal symmetry, the values of the RPV couplings  are determined by the horizontal charges of the superfields:
\beq
(\bar\mu_i, \lambda_{ijk},\lambda'_{ijk},\lambda''_{ijk})\sim\eps^{-r}(m\eps^{L_i+\phi_u}, \eps^{L_i+L_j+\ell_k},\eps^{L_i+Q_j+d_k}, \eps^{u_i+d_j+d_k}  ).
\eeqn
Taking the ratio of two couplings, the resulting textures are uniquely determined:
\beqa
\frac{\bar\mu_i}{\bar\mu_3}=\eps^{L_{i3}},\qquad \frac{\lambda_{ijk}}{\lambda_{233}}=\eps^{L_{i2}+L_{j3}+\ell_{k3}},\qquad \frac{\lambda'_{ijk}}{\lambda'_{333}}=\eps^{L_{i3}+Q_{j3}+d_{k3}},\qquad \frac{\lambda''_{ijk}}{\lambda''_{323}}=\eps^{u_{i3}+d_{j2}+d_{k3}},
\nn
\eeqan
where we can define the largest couplings $\lambda_{233}=\lambda'_{333}=\eps^{n_{LNV}}$, with $n_{LNV}=L_2+Q_3+d_3-r$, and $ \lambda''_{323}=\eps^{n_{BNV}}$,  with $ n_{BNV}=u_3+d_2+d_3-r$. The coefficients $n_{LNV}$ and $n_{BNV}$ are undetermined, and will dictate the phenomenology of the theory:
\begin{enumerate}
\item If they are fractional, all the RPV operators are forbidden;  missing energy LHC searches  apply and the weak scale is generically fine-tuned to a degree of  $\OO(1\%)$.

\item If they are both integers, $B$ and $L$ are not conserved; the leading constraint from upper limits on the proton lifetime \cite{Barbier:2004ez} is:
\beq
p\to K^+\bar\nu:\ |\lambda'_{i2k}{\lambda''}^*_{11k}|\lesssim 3\times 10^{-27} \left(\frac{m_{\tilde d_{kR}}}{100\text{ GeV}}\right)^2\\
\eeqn
For $k=3$, substituting the expressions of the couplings, we have:
\beq\label{proton}
|\lambda'_{i23}{\lambda''}^*_{113}|=\eps^{n_{LNV}+n_{BNV}+8}\lesssim \eps^{41}( m_{\tilde b_{R}}/100\text{ GeV})^2.\qquad
\eeqn
This is possible if $n_{BNV}$ and $n_{LNV}$ are 17 or higher. In section \ref{pheno}, we will see that the  couplings are tiny  and either give missing energy events or heavy stable particles. In both cases, generic limits for the sparticles masses go up to and above 1 TeV, so that this scenario does not help with low-energy SUSY.

\item If only $n_{BNV}$ is fractional, $B$ is conserved and lepton violation is allowed. This  gives rise to collider signatures with multiple leptons and LHC searches put limits near or above a TeV for a stop LSP. Again, this is not satsifactory for low-energy SUSY.

\item If $n_{LNV}$ is fractional, the only RPV operator is $\bar u\bar d\bar d$, which lets a stop LSP decay to jets and little missing energy.\footnote{Some missing energy will arise from the decays of top quarks.} Experimental searches typically involve multi-jets (6 to 10), or same-sign dileptons coming from top decays.
\end{enumerate}
We will now consider the last scenario and put bounds on the magnitude of $\lambda''$. For semi-integer $n_{LNV}$, the $L$-violating operators are forbidden, but Weinberg's neutrino mass operator is allowed, thus providing correct order-of-magnitude estimates for the neutrino masses.

\section{Phenomenological constraints}\label{pheno}
First, let us explain why arbitrarily small RPV coefficients are excluded at the LHC: in this case, the LSP does not decay in the detector, and either exits the detector as missing energy (if neutral), or hadronizes into an R-hadron (if colored). CMS currently excludes $R$-hadrons formed by a stop LSP up to 850 GeV \cite{CMS:hwa}. As
\beq
\Gamma(\tilde t\to d_id_j)=\frac{m_{\tilde t}}{8\pi}\sin^2\theta_{\tilde t}|\lambda_{3ij}''|^2,
\eeqn
for $\lambda_{323}''\lesssim \eps^{13}$, the decay length is $c\tau>1$m and the stop would hadronize. Then, a stop LSP lighter than 850 GeV has to decay in the detector and it can do so only if $\lambda''_{323}\gtrsim \eps^{13}\sim10^{-9}$. The same argument forbids a light stop decaying through a tiny coefficient in front of the operator $LQ\bar d$. Because the RPV couplings cannot be too small, $B$ and $L$ cannot be violated at the same time.

We have just presented a {\it lower} limit on the $\lambda_{323}''$; from low-energy nucleon stability, we can estimate an upper limit; for a non-R horizontal symmetry, the stronger limit comes from the neutron decay $n\to \Xi$:
\beq
|\lambda''_{112}|\lesssim 10^{-8.5}\left(\frac{m_{\tilde g}}{100\text{ GeV}}\right)^{1/2}\left( \frac{m_{\tilde s_{R}}}{100\text{ GeV}}\right)^2 \left(\frac{10^{32}yr}{\tau_{NN}}\right)^{1/4}\left(\frac{10^{-6}\text{ GeV}^6}{\langle{\bar N}|{ududss}|{\Xi}\rangle}\right)^{1/2}\nn,
\eeqn
This corresponds to $n_{BNV}\gtrsim 3$. Therefore, if low-energy supersymmetry is hidden at the LHC by baryonic \rpa\, violation, a horizontal symmetry predicts a definite texture for the couplings, and the largest couplings lays between $10^{-9}$ and $10^{-3}$.

Let us mention how low-energy SUSY is reconciled with meson mixing and the Higgs mass measurement, which both seem to point to a higher SUSY scale. First, an horizontal symmetry is naturally able to produce quark-squark alignment, in which the quark and squark mass matrices are diagonal in the same basis in which the gluino interactions are diagonal. This way, FCNCs contributing to $K-\bar K$ (and other mesons) mixing are suppressed. Second, in order to get a 126 GeV Higgs with light stops, additional tree-level contributions are needed; in \cite{Monteux:2013mna}, we considered a singlet field $N$, with a NMSSM coupling $\lambda N\phi_u \phi_d$.

\section{Conclusions}\label{conclusions}
Limits from direct stop decays in the baryonic RPV scenario are around 100 GeV and they are still coming from LEP and Tevatron.\footnote{See B.Tweedie's contribution in this conference \cite{Bai:2013xla} for a search strategy to update those limits at the LHC.} For gluinos, the situation is better: in \cite{ATLAS:2013tma}  two same-sign leptons in the final state are searched for, as a signature of two gluinos decaying as $\tilde g\to \bar t\tilde t\to\bar t bs$. The 95\% confidence level limit on the gluino mass is $m_{\tilde g}>890 $ GeV. In \cite{ATLAS-CONF-2013-091}, gluinos decaying to multiple jets were investigated, and the 95\%CL limits are at 874 GeV (these limits depend on the gluino branching ratios, which in our model can be computed). As mentioned above, the stops cannot be too much lighter than the gluinos, or we would incur in another fine tuning.

We have seen that models that explain the flavor structure of the SM naturally give hierarchical RPV couplings, suppressing those involving the first two generations, and that \rpa\, seems a superfluous assumption when looking for supersymmetric particles. If supersymmetry is to be found at LHC14, it will be stimulating to see which incarnation is reflected in the real world.

\Acknowledgments
I am grateful to Michael Dine for support and feedback during the duration of this project.

\end{document}